# Origin of Pseudogap in High-T$_c$ Superconductors

J. K. Srivastava*

*Tata Institute of Fundamental Research, Mumbai-400005, India*

(7 April 2005; revised (v3) 19 August 2005)

In an earlier paper [1] we have proposed a paired cluster (PC) model for high T$_c$ superconductivity mechanism; T$_c$=critical temperature. In this paper we show that this model is able to explain the pseudogap origin and other gap related properties of high T$_c$ superconductors (cuprates). According to the PC model, singlet coupled magnetic cluster pairs are present in the cuprates, in the material's otherwise paramagnetic state below a temperature T$_{CF}$ (T$_{CF}$=cluster formation temperature), where an ionic spin of cluster 1 (spin 1) forms a singlet pair with a corresponding ionic spin of cluster 2 (spin 2). The conducting electrons (CEs) for temperatures T≥T$_c$ and both the CEs and the Cooper pairs (CPs) for T<T$_c$ interact with the singlet coupled ion pairs by a process described in [1] and this interaction enhances the CE energy, E$_{el}$, by ΔE$_{el}$ and the CP energy, E$_{CP}$, by ΔE$_{CP}$ causing a redistribution of the filled electronic density of states (DOS). Due to this a pseudogap appears in the electronic DOS at the Fermi surface, for T$_{CF}$ ≥ T ≥ T$_c$, with d-wave symmetry which, slightly modified by ΔE$_{CP}$ enhancement, superimposes over the BCS superconducting state (SS) energy gap for T<T$_c$ resulting in (i) a mixed s-, d- wave symmetry for the observed below T$_c$ energy gap if one assumes the BCS energy gap to have anisotropic s-wave symmetry for cuprate crystal lattice (high T$_c$, anisotropic, almost no magnetic pair breaking), (ii) nondisappearance of the gap at T$_c$ on heating and almost temperature independence of the gap width, (iii) presence of states in the gap and (iv) several other gap behaviour related properties, like the absence of NMR spin relaxation rate coherence peak, which give impression of a non-BCS, nonphononic cuprate superconductivity with conducting pairs distinctly different from BCS CPs.

## I. INTRODUCTION

A large number of experimental and theoretical works have appeared in the literature concerning pseudogap presence in cuprates' normal state [1-15] and it has been suggested that a competent high T$_c$ superconductivity theory should be able to explain the pseudogap's origin and nature [9, 11, 12, 14]; T$_c$=critical temperature. The pseudogap has been observed in overdoped, underdoped and optimally doped superconductors, its presence is independent of the number of Cu-O layers in the cuprate unit cell and similar gap behaviour has been found for the electron doped and the hole doped superconductors [3, 4, 6, 7, 10, 12]. It has d-wave symmetry, develops, as one cools the lattice, in the superconductor's normal state at a temperature much above T$_c$ and remains present upto T$_c$ [2-9, 12, 14]. Below T$_c$ the observed superconducting state (SS) energy gap properties are abnormal like almost temperature (T) independent gap width, presence of states in the gap, gap's nondisappearance at T$_c$, unconventional gap symmetry, which could be a mixed s-, d- wave or anisotropic s-wave or d- wave symmetry, etc. [5, 6, 8, 12, 14, 16-35]. Several theories exist for pseudogap's origin [1, 2, 9, 11, 13, 15, 36-39], but they are found wanting in one respect or the other [1, 2, 5, 6, 11, 12, 14, 40]. In this paper we show that the paired cluster (PC) model of high T$_c$ superconductivity, developed earlier by us [1], is able to explain the pseudogap's origin and nature, the above described abnormal SS energy gap properties and also several other experimental results, like the uncommon temperature variation of NMR spin relaxation rate [41], which are related to the pseudogap and SS energy gap behaviours.

## II. ORIGIN AND CONSEQUENCES OF PSEUDOGAP

According to the PC model [1], as the magnetically frustrated cuprate lattice cools magnetic clusters are formed at a temperature T$_{CF}$ much above T$_c$; T$_{CF}$=cluster formation temperature. Different experimental techniques may sense T$_{CF}$, i.e. the clusters' presence, at somewhat different temperatures depending on their characteristic measuring times and the nature of their interaction with the clusters. Thus in practice T$_{CF}$ is either the temperature at which the clusters are formed, if the technique can sense the clusters immediately on formation, or the temperature at which the clusters are felt by the measuring technique. The clusters formed at $T_{CF}$ exist in pairs [1]. The two pair partners are interpenetrating and an ionic spin of one cluster (spin 1) forms a singlet pair with a corresponding ionic spin of its partner cluster (spin 2). As discussed in [1], for $T_c \leq T \leq T_{CF}$ the conducting electrons (CEs) interact with the singlet coupled ion pairs in the cluster and the cluster boundaries by a process described in [1]; T=temperature. This interaction enhances the CE energy, $E_{el}$, to $E_{el} + \Delta E_{el}$ and the lattice Debye temperature, $\theta_D$, to $\theta_D + \Delta\theta_D$. However for a nonzero $\Delta E_{el}$ (or $\Delta\theta_D$) a Weiss field, $H_W$, is required to exist at the ion pair site. Such a Weiss field is present for the cluster ions (CIs) since $T_{CF}$ = cluster $T_C$ (Curie temperature ) but is absent for the cluster boundary ions (CBIs) which are in the paramagnetic state at higher temperatures, $T_c \lesssim T \leq T_{CF}$. Thus for higher T only CIs contribute to $\Delta E_{el}$ and $\Delta\theta_D$ enhancements. At lower temperatures, $T \lesssim T_c$, when the CBIs freeze, which happens to be the case as is discussed later on, in spin glass (SG) state ($T < T_{SG}$, their SG temperature) [1], a nonzero $H_W$ develops at their site which enables them to contribute to $\Delta E_{el}$ and $\Delta\theta_D$. However at any T, unlike the CIs' $H_W$, which has an unique value given by the cluster magnetisation Brillouin function temperature dependence, the CBIs' $H_W$ has a distribution, ranging from zero (for certain ions) to a maximum value (for certain other ions), since in the SG state such a $H_W$ distribution is known to exist [42]. We will mention more about this at an appropriate place. As has been discussed in [1], $\Delta E_{el}$ and $\Delta\theta_D$ enhancements cause $T_c$ increase by affecting several $T_c$ influencing parameters.

For $0 \leq T \leq T_c$, Cooper pairs (CPs) exist and also those CEs which have not formed CPs. In this T range therefore both the CEs and CPs interact with the singlet coupled cluster and cluster boundary ion pairs due to which $E_{el}$ is increased to $E_{el} + \Delta E_{el}$, the CP energy, $E_{CP}$, to $E_{CP} + \Delta E_{CP}$ and the lattice $\theta_D$ to $\theta_D + \Delta\theta_D$. However the $\Delta\theta_D$ for $T \leq T_c$ is much larger than the $\Delta\theta_D$ for $T_c \leq T \leq T_{CF}$ [1]. We will concentrate below on $\Delta E_{el}$ and $\Delta E_{CP}$ enhancement effects.

The effect of above mentioned $\Delta E_{el}$ and $\Delta E_{CP}$ enhancements, which may also be called as $\Delta E_{el}$ and $\Delta E_{CP}$ scatterings, is to cause a redistribution of the filled electronic density of states (DOS), $D_f(E_{el})$, at any temperature below $T_{CF}$. This gives rise to a pseudogap in the electronic DOS distribution (redistributed $D_f(E_{el})$ vs. $E_{el}$ curve) at $T_{CF}$ which persists at lower temperatures and gets superimposed over the SS energy gap (BCS energy gap) below $T_c$. The superimposition is responsible for the abnormal SS energy gap properties. Also since $T_{CF} = T^*$ (pseudogap temperature), different experimental techniques may sense $T^*$ (i.e. pseudogap formation), like $T_{CF}$, at somewhat different temperatures. We describe the details below for different temperature ranges.

### (i) $T_c \leq T \leq T_{CF}$

In this temperature range only $\Delta E_{el}$ scattering is present as CPs do not exist. For obtaining the redistributed $D_f(E_{el})$ vs. $E_{el}$ curve the quantities needed are, the electronic DOS vs. $E_{el}$ curve which would have existed if no $\Delta E_{el}$ enhancement effect was present, T, $E_F$ (Fermi energy), $\Delta E_{el}$ and $N_P$, the percentage of $CE_S$ for which $\Delta E_{el}$ enhancement occurs.

The calculation of $\Delta E_{el}$ is given in [1] and it ($\Delta E_{el}$) depends on T and $E_{el}$ through the magnetic fields $H_W$, $H_{dip}$, $H_{CE}$ (and also $H_{CP}$ for $T < T_c$) and relaxation times $\tau'$ and $\tau$, all defined in [1]. The $H_W$, to a good approximation, can be obtained by assuming a Brillouin function temperature dependence for the cluster magnetisation; this is since $H_W$ does not exist for CBIs at high T. However the calculation of $\tau'$, $\tau$ is generally not possible [1] and only an order of magnitude estimate can be made for them. Using such estimates [1], $\Delta E_{el}$ is found to have an oscillatory dependence on T. For a given $E_{el}$, as T decreases below $T_{CF}$ $\Delta E_{el}$ first increases upto a certain temperature, then becomes almost constant over a temperature range below which it decreases. Such a behaviour arises owing to the exponential dependence of the ionic level Boltzmann population on $H_W/T$ and the $H_W$'s Brillouin function T dependence due to which $\Delta H_W/\Delta T$ decreases as T decreases, tending to zero as $T \to 0$. Approximately $T/T_C (\equiv T/T_{CF}) \sim 0.5$ could be taken as the temperature at which the $\Delta E_{el}$ is maximum.

At a given T, $\Delta E_{el}$ depends on $E_{el}$. The nature of this dependence is governed by the $E_{el}$ dependence of $\tau$ which can not be exactly calculated [1] and therefore various possibilities are to be examined. Thus $\Delta E_{el}$ vs. $E_{el}$ could be flat ($\Delta E_{el}$ independent of $E_{el}$) or linear or exponential depending on the magnitude of $\tau$ and whether $\tau$ vs. $E_{el}$ is flat or linear or exponential. Physically, an exponential dependence of $\tau$ on $E_{el}$ is more probable as is seen for the paramagnetic ion concentration dependence of the ionic spin-spin relaxation time [43].

The $N_P$ is another needed quantity. At high T when CBIs do not contribute to $\Delta E_{el}$ enhancement, $N_P \sim 50\%$ if one assumes almost equal number for CIs and CBIs i.e. almost equal total occupied volumes for the clusters and the cluster boundaries in the lattice. At low T, CIs' contribution to $\Delta E_{el}$, and $\Delta E_{CP}$, diminishes and finally vanishes. However CBIs contribute to $\Delta E_{el}$, and $\Delta E_{CP}$, at these temperatures. But owing to $H_W$ distribution only few (a fraction of) CBIs, which have right magnitude $H_W$ at the working T, contribute. Thus $N_P$ is small and has no systematic T variation. $N_P \sim 50\%$ is also an approximation since the cluster boundary volume is not known [1]. It is therefore better to use the experimental results as guidelines for estimating $N_P$ when comparing experiment and theory. This is true for other parameters also.

Thus for getting the theoretical results, the parameter values used have been obtained by using both the theoretical considerations [1] and the experimental results [6, 20-22, 26, 44] as guidelines. The two guidelines have been found to give consistent estimates. Further though the results discussed below are for typical parameter values, calculations have been done for other parameter values also and the results obtained are similar in nature. This is mentioned at appropriate places describing the parameters used and the results obtained.

Fig. 1 shows a typical result where the electronic DOS, $D(E_{el})$, is plotted against $E_{el}$. The dotted curve is the total (filled plus empty) DOS, $D_t(E_{el})$, (quadratic, free electron approximation [45], Appendix), the dashed curve is the density of filled states, $D_f(E_{el})$, which would have existed at temperature T if there was no $\Delta E_{el}$ enhancement present, the full line curve is the density of filled states redistributed, i.e. redistributed $D_f(E_{el})$ or $D_{fr}(E_{el})$, due to $\Delta E_{el}$ enhancement effect and in Fig. 1(a) the dash-dot curves a, b are the same as the full line curve there but have been obtained for different $N_P$ values; $N(E_{el})$, the number of CEs at energy $E_{el}$, $= 2 D(E_{el})$. Fig. 1(a) shows the results for $T \sim T_{CF}$ case. For $YBa_2Cu_3O_7$, $T_{CF} \sim 220K$ and $E_F \sim 310$ meV [1]. For other cuprates also $T_{CF}$ and $E_F$ are of similar order. We therefore assume these values for the present calculations. However we have done calculations for the higher and lower values of $T_{CF}$ and $E_F$ also and the results obtained are similar to the results given here. For Fig. 1(a), the various parameters used in the calculation are, $T = 200K$ ($T \sim T_{CF}$), $E_F = 310$ meV, $\Delta E_{el}(E_F)$, the value of $\Delta E_e$ at $E_{el} = E_F$, $= 300$ meV, an exponential dependence, of the form given below, is assumed for $\Delta E_{el}$ on $E_{el}$ and $N_P = 50\%$ (full line curve), 40% (curve a) and 60% (curve b). The $\Delta E_{el}$ vs. $E_{el}$ form is: $\Delta E_{el} = (\Delta E_{el})_0 [1-\exp(-\alpha E_{el})]$, where $(\Delta E_{el})_0$, $\sim E_{el}(E_F)$, $= 300$ meV, $\alpha = 0.7$ (meV)$^{-1}$ i.e. a quick rising, fast saturating $\Delta E_{el}$ variation with $E_{el}$ is assumed. The $\Delta E_{el}(E_F)$ given above matches with the theoretical estimate [1] and with that deduced from the experimental results [6]. Owing to Pauli principle, the CE energy can be enhanced from $E_{el}$ to $E_{el} + \Delta E_{el}$ only if the energy state at $E_{el} + \Delta E_{el}$ is either partially or completely empty. Due to this the full line curve (Fig. 1(a)) does not start from the origin but merges with the dashed curve at $E_{el} \sim 3$ meV; this effect can be seen more clearly in the below discussed Fig. 1(b). Similar behaviour exists for the curves a, b (Fig. 1(a)) also but is not shown in the figure for clarity near $E_{el} \sim 0$ region. If instead of an exponential $E_{el}$ dependence, an $E_{el}$ independence is assumed for $\Delta E_{el}$, then the full line, a, b curves start from the origin but the other results, like their nature etc., are similar. Similarly, for a linear $E_{el}$ dependence of $\Delta E_{el}$, though the full line, a, b, curves merge with the dashed curve at $E_{el} >> 3$ meV, the other results obtained are similar to those shown in the figure (Fig. 1(a)). The full line, a, b, curves are calculated by adding the scattered CEs, at any energy state, to the CEs which had remained there after the scattering from that state had taken place, if the state was not completely empty (Appendix). A comparison of the full line, a, b curves shows the $N_p$ value's effect.

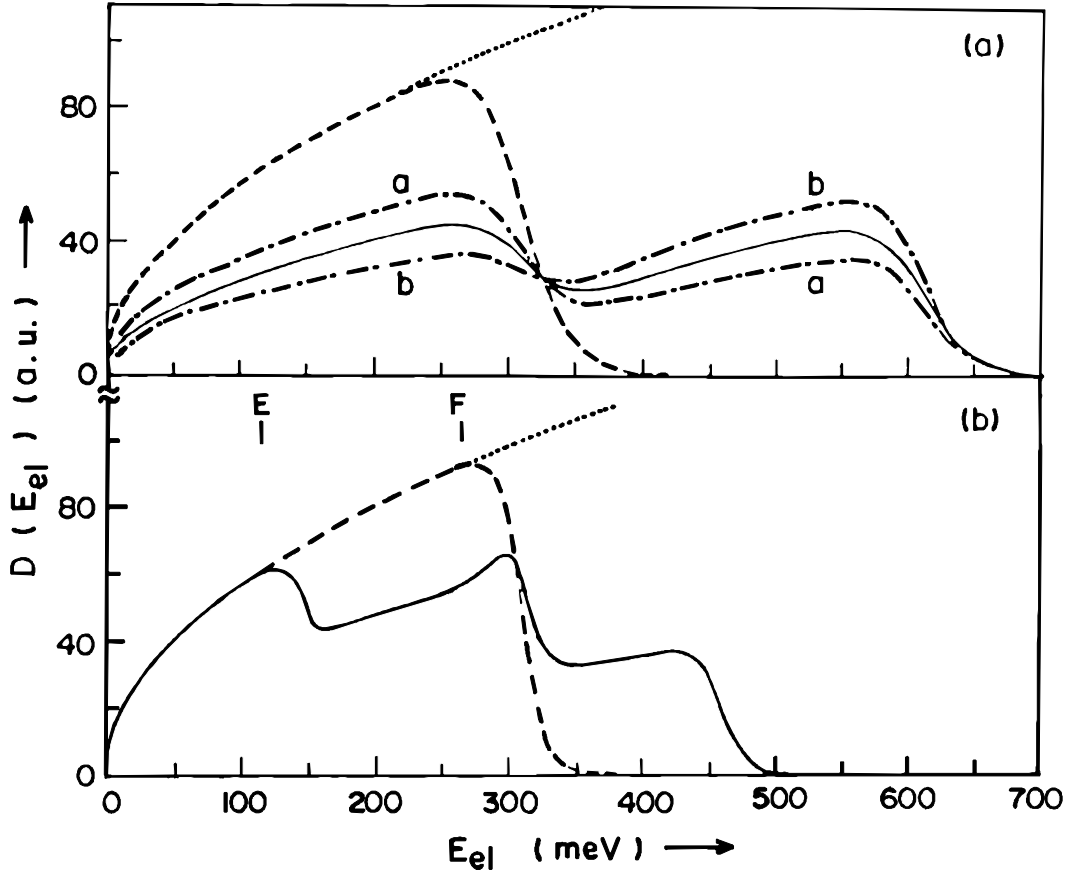

Fig. 1. Dependence of the electronic density of states, $D(E_{el})$, on electrons' energy, $E_{el}$, for $T > T_c$; $T_c$ = critical temperature, a.u. = arbitrary unit. Details are described in the text.

As mentioned above, for calculating the Fig. 1(a) $D_{fr}(E_{el})$ curves we have used the dotted $D_t(E_{el})$ curve. However we have done calculations for other types of $D_t(E_{el})$ curve also which instead of increasing with $E_{el}$ in the $E_{el} \sim E_F$ region, as in Fig. 1(a), either remain flat ($E_{el}$ independent) or decrease with increasing $E_{el}$. The results obtained in all the cases are similar to what is shown in Fig. 1(a). A pseudo-energy gap (pseudogap) is clearly seen in the $D_{fr}(E_{el})$ vs. $E_{el}$ distribution in Fig. 1(a) at the Fermi surface. Similar pseudogap is seen in Fig. 1(b) also where all descriptions, and parameter values, are same as those of Fig. 1(a) except T = 100K, $\Delta E_{el}(E_F)$ = 150 meV and $N_P$ = 40%. Thus Fig. 1(b) describes a situation where $T_c < T < T_{CF}$. The full line curve in Fig. 1(b) meets the dashed curve at much higher $E_{el}$ than the curves of Fig. 1(a) due to a smaller $\Delta E_{el}$ owing to which the Pauli principle does not allow CEs from the states below the energy state marked E to be scattered since the empty (partially or fully) states occur from the energy state marked F onwards in the $D_t(E_{el})$ vs. $E_{el}$ distribution.

Thus we see that a pseudogap appears in the $D_{fr}(E_{el})$ vs. $E_{el}$ distribution (Fig. 1) for $T \leq T_{CF}$. This pseudogap persists for $T \leq T_c$ also since $\Delta E_{el}$ scattering is present below $T_c$ too where, in addition, $\Delta E_{CP}$ scattering also occurs whose effect we will discuss in a later section. The $D_{fr}(E_{el})$ vs. $E_{el}$ distribution of Fig. 1 when translated into the tunneling conductance curves [46, 47] shows agreement with the experimental results [6] like the shape, size, location and T dependence of the pseudogap. Since the pseudogap arises due to the CE ($\Delta E_{el}$) scattering, it has d-symmetry (actually mixed d-, p- ($d_{x2-y2}$-, $p_x$-, $p_y$-) symmetry which experimentally can not be distinguished from a d- symmetry) in the normal state. This is shown by the experiments [5, 6, 8, 9, 12, 14, 37]. Below $T_c$, $\Delta E_{CP}$ scattering and SS energy gap are also present and this situation gap symmetry we will describe in the later section. Our results are also consistent with various other experiments which conclude that the pseudogap arises in the charge excitation spectrum [7, 11, 12, 14].

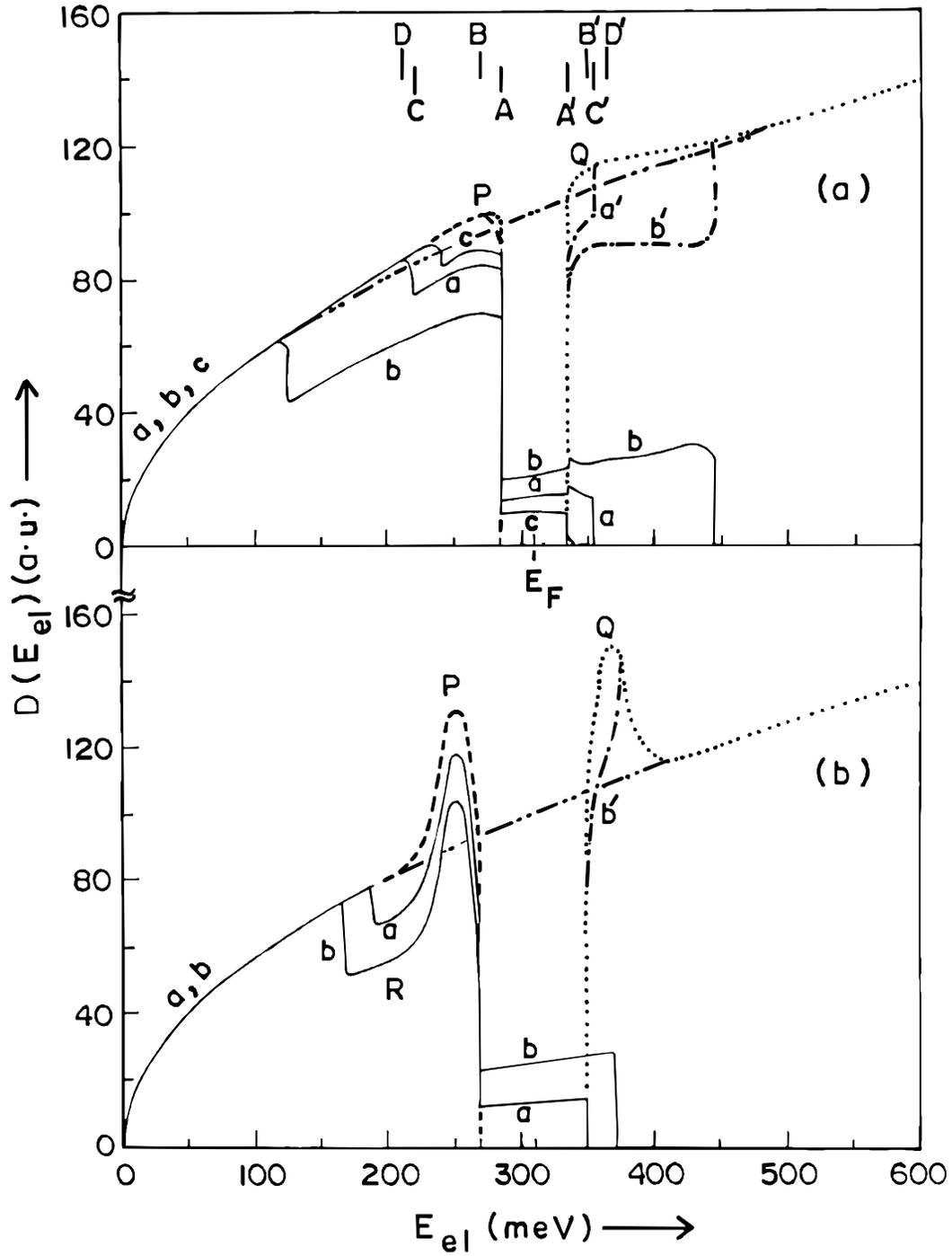

Fig. 2. Dependence of the electronic density of states, D($E_{el}$), on electrons' energy, $E_{el}$, for $T \lesssim T_c$; $T_c$ = critical temperature, a.u. = arbitrary unit. Details are described in the text.

**(ii) $0 \leq T \leq T_c$**

As mentioned above, below $T_c$ the SS energy gap (BCS energy gap [1, 45]) appears which is present alongwith the $\Delta E_{el}$ scattering and additional $\Delta E_{CP}$ scattering (enhancement) near $T_c$; this is since, as will be discussed later, $\Delta E_{CP} \sim 0$ for T away from $T_c$. This ($0 \leq T \leq T_c$) case is more complex and the results obtained are summarised in Figs. 2, 3. We first discuss the Fig. 2 results where $T \sim T_c$ (Fig. 2(a)) and $T < T_c$ (Fig. 2(b)). In Fig. 2(a), 2(b), $E_{el}$ variation is shown for $D_t(E_{el})$ (dotted curve), $D_f(E_{el})$ (dashed curve), $D_{fr}(E_{el})$ (full line curve) and redistributed $D_e(E_{el})$, the redistributed density of empty states (shown only for above the BCS energy gap region), (dash-dot curve). The dash-double dot curve is an extension of the $D_t(E_{el})$ curve which would have existed if there were no SS peaks and energy gap present. In a tunneling conductance measurement, the SS energy gap at T, $2\Delta(T)$, is measured either by the SS peaks' (P, Q in Fig. 2(a)) separation or by A, A′ (Fig. 2(a)), the points at which the dash-double dot curve intersects the SS energy gap edges, separation [6, 44]. The curve a′(b′) is obtained by subtracting the unoccupied (empty) side ($E_{el} > E_F + \Delta(T)$, $\Delta(T)$ = SS energy gap width/2) portion of the curve a(b) from the $D_t(E_{el})$ curve. Writing $D_t(E_{el}) = D_f(E_{el}) + D_e(E_{el}) = D_{fr}(E_{el}) + D_{er}(E_{el})$, curve a′(b′) is actually the empty side portion of the $D_{er}(E_{el})$ (i.e. redistributed $D_e(E_{el})$) curve which gets merged with the dotted curve.

As before, various Fig. 2 parameters have been chosen using theoretical and experimental [1, 6] considerations. In Fig. 2(a), T = 75K, $E_F$ = 310 meV, $2\Delta(T)$ = 50 meV (i. e. T spread of $D_f(E_{el})$ curve's tail portion $\sim 2\Delta(T)$) and, as before, an exponential dependence is assumed for $\Delta E_{el}$ on $E_{el}$. This (exponential dependence) assumption has been used in Fig. 2(b), 3 calculations also. For the curve a, $\Delta E_{el}(E_F)$ = 70 meV and $N_P$ = 15%, for curve b, $\Delta E_{el}(E_F)$ = 165 meV and $N_P$ = 30%, and for the curve c, $\Delta E_{el}(E_F)$ = 50 meV and $N_P$ = 10%. As mentioned before, CPs also exist at this T and calculations show that for them $\Delta E_{CP} \sim \Delta E_{el}(E_F)$ for all $E_{el}$ [1] and, for obvious reasons, they have the same percentage scattering as the CEs (i.e. $(N_P)_{CP} = (N_P)_{CE}$). We have done Fig. 2(a) calculations assuming these but even for somewhat different values of $\Delta E_{CP}$ and $(N_P)_{CP}$, the results obtained are similar mainly because the CPs are much smaller in number than the CEs [1] and so have only small influence on the results. Also in these calculations, as well as in Fig. 2(b), 3 calculations, it has been assumed that CEs can have scattered states in the SS energy gap as happens in the case of the inelastic electron scattering and the proximity effect [46, 48-51]. However even without this assumption similar results are obtained when the redistributed DOS curves are translated into the tunneling conductance curves [6]. This happens due to the existence of CEs' scattered states outside the SS energy gap. The assumption of nonzero CE scattering in the SS energy gap, however, looks justified as the experimental results too support the presence of states in the SS energy gap [6, 20, 21, 26]. It has also been assumed that CPs do not get scattered in the SS gap. Though even without this assumption the results similar to those of Figs. 2, 3 are obtained, mainly owing to the much smaller percentage of CPs, the assumption is physically justified since the interactions causing $\Delta E_{el}$, $\Delta E_{CP}$ scatterings are not pair breaking [1]; they rather enhance $T_c$, $\Delta(0)$, the T =0K SS energy gap width/2 [1]. As has been explained before at these temperatures (T $\sim$ 75K and below, when $T_{CF} \sim$ 220K [1]) CIs' contribution to $\Delta E_{el}$, $\Delta E_{CP}$ decreases and CBIs' contribution becomes significant due to their SG freezing. However CBIs have $H_W$ distribution and a relatively smaller maximum $H_W$. The $N_P$ and $\Delta E_{el}$ are therefore smaller at these temperatures.

Fig. 2(a) results can now be understood. The curves a, a′ show that in this case due to the filled and empty states' redistribution owing to $\Delta E_{el}$, $\Delta E_{CP}$ scatterings, both, the filled side and the empty side, SS peaks, P and Q, have got modified in location, shape and size. Their peak positions have shifted from B, B′ to D, D′ showing P, Q separation enhancement. Similarly A, A′ separation has increased to C, C′ separation. Since, as mentioned before, $\Delta(T)$ is measured by these separations [6, 44], the effect of these separation enhancements, which are relatively larger at higher T due to large $\Delta E_{el}$, $\Delta E_{CP}$, is to cause an increase in the value of $2\Delta(T)$. This, to a good extent, nullifies the $\Delta(T)$'s BCS increasing - T decrease. As a result the observed $\Delta(T)$ shows weak T dependence [6]. The $\Delta(T)$'s increasing- T decrease is also partially compensated by the following effect, more effective at higher T, of the $\Delta E_{el}$, $\Delta E_{CP}$ scattering induced DOS redistribution. If 2f(T) is the fraction of broken CP CEs (quasiparticles), i.e. broken CPs' fraction, near the SS energy gap edge at T (2f(0) = 0, 2f($T_c$) = 1), then [47, 52] approximately $\Delta(T) \propto [1-2f(T)]$. Since due to the DOS redistribution, f(T) is reduced near the gap edge, as $D_{fr}(E_{el}) < D_f(E_{el})$, near the filled side gap edge, shows (Figs. 2, 3), $\Delta(T)$ gets enhanced. The f(T) decrease near the gap edge is a consequence of the fact that alongwith the normal CEs, broken CP CEs also get scattered due to the $\Delta E_{el}$ scattering. Experimentally [6] the SS energy gap observed below $T_c$ does not vanish at $T_c$ whereas BCS $\Delta(T) = 0$ at $T_c$. This happens because at $T_c$ the earlier described $\Delta E_{el}$ scattering induced normal state pseudogap is present. Since like the pseudogap just above $T_c$ the SS energy gap just below $T_c$ has $\Delta E_{el}$ scattering induced states in the gap, there is no conspicuous change in the gap's nature (width, shape, size, location) at $T_c$. The gap vanishes only at $T_{CF}$. These results agree with the experiments [6].

The SS energy gap symmetry can also be understood now. The observed SS gap is the BCS energy gap modified by the $\Delta E_{el}$, $\Delta E_{CP}$ scattering induced effects (pseudogap effect). As discussed before, the $\Delta E_{el}$ scattering induced pseudogap has a d-wave symmetry. The BCS gap in principle can have any symmetry [1, 18, 33, 35]. However physically for the cuprates (high $T_c$, anisotropic crystal structure), an anisotropic s-wave symmetry is more likely for the BCS energy gap since non-s wave superconductors are known to have low $T_c$ [53] and s-wave CP coupling is more stable against magnetic perturbations (paramagnetic ions, electric current, magnetic field). Thus the observed SS energy gap below $T_c$ is expected to have a mixed anisotropic s-, d- wave symmetry. Experimentally it is not possible to distinguish easily between an anisotropic s-wave or d-wave or mixed s-, d- wave symmetry and therefore experimentalists favour one symmetry or the other [1, 5, 6, 8, 12, 14, 16-35].

The curves b, b′ (Fig. 2(a)) show that in this case both, the P and Q, peaks have disappeared as a result of the DOS redistribution. Thus eventhough we are below $T_c$, experiments, like the tunneling experiments, may indicate a normal state, with a pseudogap, for the system suggesting a lower $T_c$ value. Also such a DOS redistribution may be one of the reasons for the NMR spin relaxation rate coherence peak's absence in cuprates since P peak's presence is necessary for the coherence peak's existence [41]. The curve c (Fig. 2(a)) shows that for this case only the peak P has got modified (height, width decreased, position shifted) and the peak Q has remained undisturbed. In this case also the measured $\Delta(T)$ will be larger than the $\Delta(T)$ which would have been obtained if no DOS redistribution existed. Such a $\Delta(T)$ enhancement, as mentioned before, partially compensates for any $\Delta(T)$'s T decrease. All these results, and the nature (height, width, shape, location) of the modified P, Q peaks and of the observed SS energy gap, agree with the experiments [6].

In Fig. 2(b), T = 50K, $E_F$ = 310meV and $2\Delta(T)$ = 80 meV (i.e. $D_f(E_{el})$ curve's tail portion's T spread < $2\Delta$). For the curve a, $\Delta E_{el}(E_F)$ = 80 meV, $N_P$ = 15% and for the curve b, $\Delta E_{el}(E_F)$ = 100 meV, $N_P$ = 30%. For both the cases, $\Delta E_{CP} \sim 0$ ( < 1 meV), both for CIs and CBIs [1], and therefore the $\Delta E_{CP}$ scattering effect is negligible. For the case of curve a, due to the DOS redistribution, owing primarily to the $\Delta E_{el}$ scattering, peak P height, width decrease, peak Q remains undisturbed and dip R (a part of the curve a) appears on the P peak side. This dip is more pronounced for the curve b where both P and Q have got modified. These results and the relative modified peaks' and dips' heights, widths, locations, shapes etc. agree with the experimental results [6]. Also for both the cases, the measured $\Delta(T)$ is greater than the non-DOS redistributed case's $\Delta(T)$. This, as before, makes $\Delta(T)$'s T dependence very weak.

We can now discuss the results of Fig. 3 where the full line curve, dotted line curve etc. have the same meaning, and the curves a′, b′, c′ are obtained in the same way, as in Fig. 2 and T << $T_c$ (Fig. 3(a)), T near $T_c$ (Fig. 3(b)). We first discuss Fig. 3(a) results where the parameters used are, T = 4.2K, $E_F$ = 310 meV, $2\Delta(T)$ = 90meV (i.e. $D_f(E_{el})$ curve's tail portion's T spread (~ 0) << $2\Delta$), $\Delta E_{CP} \sim 0$ (< 1 meV), $\Delta E_{el}(E_F)$ = 105 meV and $N_P$ = 15%. As before, the DOS redistribution's effect is to enhance $2\Delta(T)$. However since $\Delta(4.2K) \sim \Delta(0)$, this enhancement means enhanced $2\Delta(0)/k_BT_c$ ratio ($k_B$ = Boltzmann's constant). Thus to some extent the large $2\Delta(0)/k_BT_c$ ratio (> BCS weak coupling limit) for cuprates can be understood from the DOS redistribution. However for accounting it fully, the effect of somewhat large $\lambda$, S′ parameters, defined in [1], is to be taken into account. The relative heights, widths, locations, shapes etc. of the dip and the peaks obtained here (Fig. 3(a)) agree with the experimental results [6].

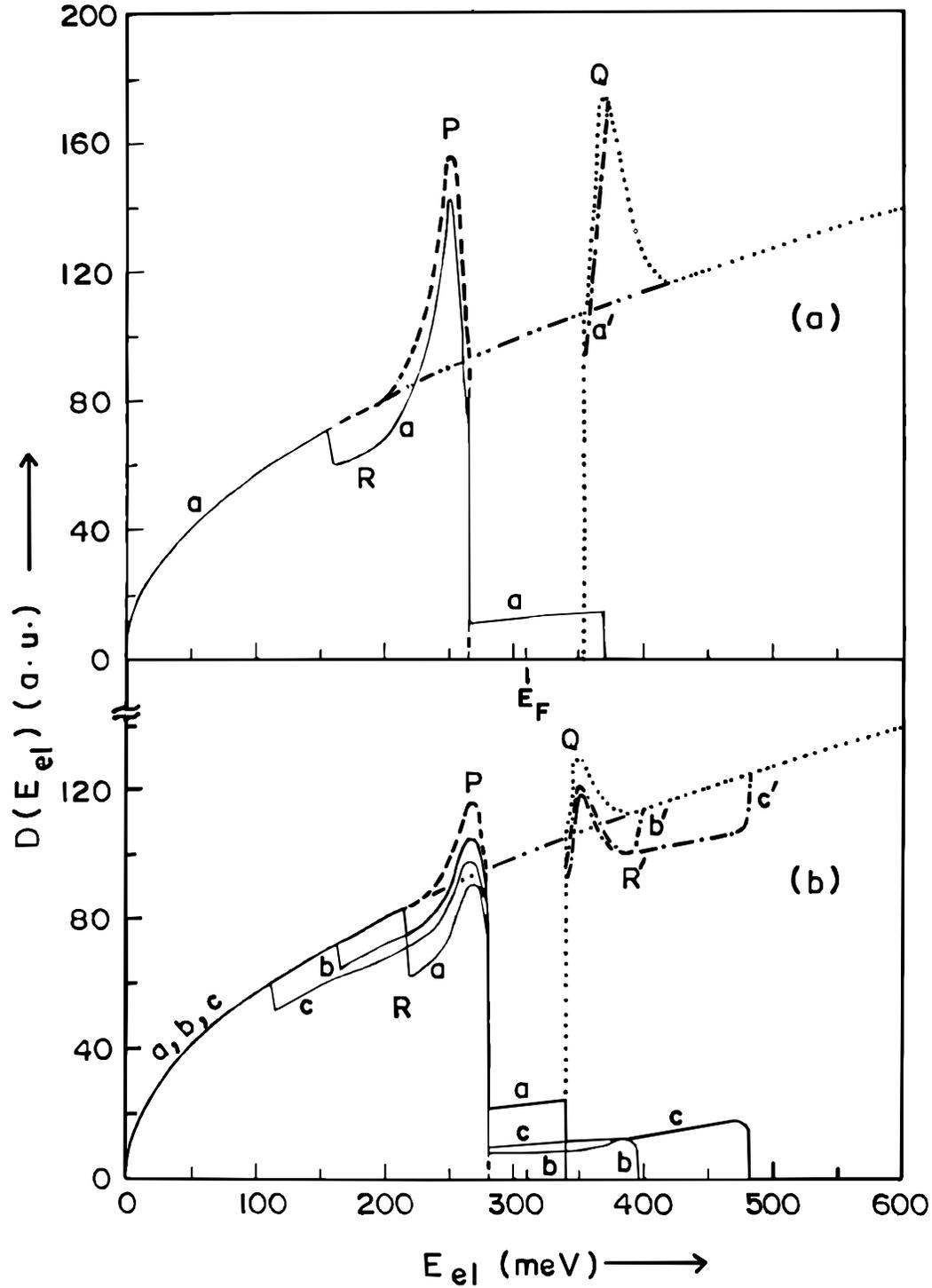

Fig. 3. Dependence of the electronic density of states, $D(E_{el})$, on electrons' energy, $E_{el}$, for $T < T_c$; $T_c$ = critical temperature, a.u. = arbitrary unit. Details are described in the text.

For Fig. 3(b), $T = 70K$, $E_F = 310$ meV, $2\Delta(T) = 60$ meV (i. e. $D_F(E_{el})$ curve's tail portion's T spread $\sim 2\Delta(T)$), $\Delta E_{CP} \sim \Delta E_{el}(E_F)$ for all $E_{el}$ and $(N_P)_{CP} = (N_P)_{CE}$. This figure (3(b)) shows some specific cases. For the curve a, $\Delta E_{el}$ ($E_F$) = 60 meV, $N_P$ ($\equiv (N_P)_{CE}$) = 25%, for the curve b $\Delta E_{el}$ ($E_F$) = 9 meV, $N_P$= 10% and for the curve c, $\Delta E_{el}$ ($E_F$) = 165 meV, $N_P$ = 15%. Whereas for the curve a's case, peak P disappears and peak Q remains undisturbed, for the case of curves b, b′ both P, Q almost symmetrically decrease and a dip R′ appears on the peak Q side. This dip is broadened for the case of curves c, c′ where the peak P has very small amplitude. These results agree with the experimental results [6]. As before these DOS redistributions too enhance $\Delta(T)$, partially compensating for $\Delta(T)$'s T decrease and making $\Delta(T)$'s T dependence weak. R, R′ having locations dependent on P, Q (gap edges) positions are experimentally seen [6, 21].

## III. CONCLUSION

In conclusion, the cuprate properties can be understood on the basis of what is described in this paper and in [1]. This is true even for those properties which have not been specifically discussed here. As an example we explain below the anomalous T behaviour of the NMR spin relaxation rate ($1/T_1$) [41]. The coherence peak's absence in $1/T_1$ vs. T, near $T_c$, has already been explained earlier. The other $1/T_1$ vs. T anomalies can be understood in a similar way. For instance, $1/T_1$ decreases with decreasing T due to lattice cooling and shows drops at $T^*$ ($\gg T_c$, ~ 220K for $YBa_2Cu_3O_7$ ), $T_c$ [41]. The drop at $T_c$ is seen in conventional (elemental, A15) superconductors also and results from the CPs' presence. Assuming same explanation for cuprates' $T_c$ drop, the drop at $T^*$ is not yet properly understood. For example one of the explanations assumes CPs' formation at $T^*$ itself [38]; CPs grow in size with decreasing T, developing overlap and phase coherence below $T_c$ to give superconductivity. Experiments do not support this explanation [1, 12, 14]. Also CPs' stability is difficult to understand without their phase coherence [1]. If one assumes too tightly bound a CP at $T^*$, then the problems of lattice distortion, lattice stability and polaron formation in cuprates come into the picture [1]. In our model, $T^* = T_{CF}$ and the lattice excessively cools at this temperature due to the interactions described in PC model [1]. This excessive cooling is responsible for the drop in $1/T_1$ at $T^*$. The drop at $T_c$ results from both the CPs' presence and the excessive lattice cooling at $T_c$ [1]. Thus the anomalies of the $1/T_1$ vs. T behaviour, like the Mössbauer f-factor vs. T anomalies [1], can be understood on the basis of the channeling r.m.s. lattice ions' vibration amplitude vs. T data [1]. Similarly $T^*$ (pseudogap temperature) vs. H can also be understood on the basis of our (PC) model. Apart from influencing the frustration parameters ($\tilde{J}_0, \tilde{J}$), H can also affect [1] α, τ, $\Delta E_{el}$, $\Delta E_{CP}$, $N_P$, electron motion (via magnetoresistance), RVB singlet coupling of paired clusters' spins [1] by tilting (aligning) the moments towards (along) $\vec{H}$, etc. Therefore $T^*$ vs. H behaviour is complex ($\vec{H}$-, system-dependent). However, a large enough $\vec{H}$ can appreciably weaken the RVB singlet coupling by, as mentioned before, tilting (aligning) the spins towards (along) it and thus making it necessary to go to a lower temperature to get the RVB singlet coupling back. Thus the pseudogap temperature ($T^*$) gets decreased with H for large H. This has been experimentally observed [54]. In the same way, the Zeeman scaling relation observed in [54] between the T~ 0K pseudogap closing field, $H_{pg}(0)$, and the pseudogap temperature $T^*$ too can be understood easily on the basis of our (PC) model according to which the RVB singlet coupling energy ($E_{RVB}$) ~ $k_B T^*$ ~ $\mu H_{pg}(0)$. Thus $\mu H_{pg}(0)$ ~ $k_B T^*$, where $\mu = g\mu_B S$ and S=1/2, g=2 for $Cu^{2+}$ ions which have orbital singlet ground state in the cuprate crystal field [1]. This is what has been observed experimentally as Zeeman energy scaling relation. Physically $H_{pg}(0)$ is large enough to align all the cluster and cluster boundary ions along $\vec{H}_{pg}(0)$ and break the RVB singlet coupling at T=0K (which closes the pseudogap). Thus $\mu H_{pg}(0)$ gives the RVB singlet coupling energy. Similarly H=0 $T^*$ (i.e. $T^*(H=0) \equiv T_{CF}$) is the large enough temperature above which RVB singlet coupling is destroyed and system becomes paramagnetic. Thus $k_B T^*$ also give the RVB singlet coupling energy. Therefore $\mu H_{pg}(0)$ ~ $k_B T^*$. This is similar to the relation $\mu H_W(0)$ ~ $k_B T_N(T_C)$ of magnetically ordered systems. Finally, the results (Figs. 1-3) given here are for three dimensional (3D) case (Appendix). However these results have been found to remain same even for a two dimensional (2D) case [1, 55]. Thus the discussions given here are valid whether a cuprate lattice has more two dimensionality (2D nature) or more three dimensionality [1, 55].

## IV. SUMMARY

We have described here some consequences of a new mechanism (PC model) of high $T_c$ superconductivity [1]. According to our cluster phase transition (CPT) model of spin glass systems [1], magnetic clusters are present in the frustrated magnetic lattices in the material's otherwise paramagnetic state below a temperature $T_{CF}$ (cluster formation temperature). In PC model, it has been envisaged that in high $T_c$ superconducting cuprate systems, which are magnetically frustrated, these magnetic clusters exist in pairs. The two pair partners are interpenetrating and an ionic spin of one cluster forms a singlet pair with a corresponding ionic spin of the partner cluster. This singlet pairing could occur due to a resonating valence bond (RVB) interaction. The conducting electrons (CEs) and the Cooper pairs (CPs), formed by BCS-Migdal-Eliasberg phonon coupling, interact with the singlet coupled ion pairs and this interaction is responsible for the $T_c$ enhancement and other properties of cuprate superconductors.

Further the above model, called paired cluster (PC) model of high $T_c$ superconductivity, is able to explain the pseudogap origin and other gap related properties of high $T_c$ superconductors (cuprates). The interaction of CEs for temperatures $T \geq T_c$ and of both the CEs and the CPs for $T < T_c$ with the singlet coupled ion pairs enhances the CE energy, $E_{el}$, by $\Delta E_{el}$ and the CP energy, $E_{CP}$, by $\Delta E_{CP}$ causing a redistribution of the filled electronic density of states (DOS). Due to this a pseudogap appears in the electronic DOS at the Fermi surface, for $T_{CF} \geq T \geq T_c$, with d-wave symmetry which, slightly modified by $\Delta E_{CP}$ enhancement, superimposes over the BCS superconducting state (SS) energy gap for $T < T_c$ resulting in (i) a mixed s-, d- wave symmetry for the observed below $T_c$ energy gap if one assumes the BCS energy gap to have anisotropic s-wave symmetry for cuprate crystal lattice (high $T_c$, anisotropic, almost no magnetic pair breaking), (ii) nondisappearance of the gap at $T_c$ on heating and almost temperature independence of the gap width, (iii) presence of states in the gap and (iv) several other gap behaviour related properties, like the absence of NMR spin relaxation rate coherence peak, which give impression of a non-BCS, nonphononic cuprate superconductivity with conducting pairs distinctly different from BCS CPs.

## APPENDIX

For three dimensional (3D) lattice, for $T \geq T_c$ $D_t(E_{el}) = A E_{el}^{1/2}$ and for $T < T_c$,

$$D_t(E_{el}) = A \operatorname{Re}[(E_F \pm \sqrt{[(E_{el} + i\Gamma) - E_F]^2 - \Delta^2})^{1/2} \{\frac{|(E_{el} + i\Gamma) - E_F|}{\sqrt{[(E_{el} + i\Gamma) - E_F]^2 - \Delta^2}}\}] \quad ,$$

where - (minus) sign before the square root, in the first round bracket, applies for $E_{el} \leq E_F$ and + sign for $E_{el} > E_F$, Re means real part, A= proportionality constant, T = working temperature, $i = (-1)^{1/2}$, $\Delta$ = BCS (superconducting state (SS)) energy gap/2, $E_F$ = Fermi energy and $\Gamma$ = broadening due to Cooper pair (CP) lifetime decay. For $T \geq T_c$, the other quantities are as follows.

$$D_f(E_{el}) = D_t(E_{el}) \times f(E_{el}), \qquad (i)$$

where $f(E_{el}) = 1/\{\exp[(E_{el} - E_F)/k_B T] + 1\}$ and $k_B$ = Boltzmann's constant. Assuming that the Pauli principle permits the above mentioned $\Delta E_{el}$ scattering, i.e. empty states are available for such a scattering, we have [1],

$$D_{fr}(E_{el}) = D_f(E_{el}) - N_P D_f(E_{el}) + N_P D_f(E_{el} - \Delta E_{el}). \qquad (ii)$$

For $T \sim T_c$ case, where $\Delta E_{CP}$ scattering is also present [1], we have for the region below the gap ($E_{el} < (E_F - \Delta)$),

$$D_{fr}(E_{el}) = D_f(E_{el}) - N_P D_f'(E_{el}) + N_P D_f'(E_{el} - \Delta E_{el}) - (N_P)_{CP} [D_f(E_{el}) - D_f'(E_{el})] + (N_P)_{CP} \times$$

$$[D_f(E_{el} - \Delta E_{el}) - D_f{'}(E_{el} - \Delta E_{el})]. \qquad (iii)$$

The third term of the above equation will be nonzero only in very few cases when some empty states are present in $D_t{'}(E_{el})$ below the gap, like, for example, Fig. 2(a) case where the temperature spread of $D_f(E_{el})$ curve's tail portion is slightly greater than the gap value. Similarly for the region inside the gap we have,

$$D_{fr}(E_{el}) = N_P\, D_f{'}(E_{el} - \Delta E_{el}) + D_f(E_{el}). \qquad (iv)$$

The second term in the above equation contributes only if there are some filled states in the gap; generally this is zero or very small. Finally, for the region above the gap,

$$D_{fr}(E_{el}) = N_P\, D_f{'}(E_{el} - \Delta E_{el}) + D_f(E_{el}) + (N_P)_{CP}\,[D_f(E_{el} - \Delta E_{el}) - D_f{'}(E_{el} - \Delta E_{el})]. \qquad (v)$$

As in Eq.(iv), the second term of Eq.(v) is zero or very small. The third term of Eq.(v) contributes only if some excited CPs are present above the gap.

For $T < T_c$, $(N_P)_{CP} \sim 0$ since $\Delta E_{CP} \sim 0$; $N_P \equiv (N_P)_{CE}$ and $(N_P)_{CP}$ = percentage (fraction) of CPs for which $\Delta E_{CP}$ enhancement occurs.

---

\* Electronic address: jks@tifr.res.in

[1]  J. K. Srivastava, Phys. Stat. Sol. (b) **210**, 159 (1998); J. K. Srivastava, in "Models and Methods of High-$T_c$ Superconductivity: Some Frontal Aspects, Vol. 1", eds. J. K. Srivastava and S. M. Rao (Nova Science, New York, 2003), p. 9; J. K. Srivastava in "Studies of high Temperature Superconductors: Advances in Research and Applications, Vol. 29", ed. A. Narlikar (Nova Science, new York, 1999), p. 133; J. K. Srivastava, cond-mat/0504245 (April 2005); cond-mat/0508292 (August 2005).
[2]  K.F. Quader and G.A. Levin, Phil. Mag. B **74**, 611 (1996).
[3]  J. W. Loram, K. A. Mirza, J. M. Wade, J. R. Cooper, and W. Y. Liang, Physica C **235-240**, 134 (1994).
[4]  D. N. Basov, T. Timusk, B. Dabrowski, and J. D. Jorgensen, Phys. Rev. B **50**, 3511 (1994).
[5]  G. V. M. Williams, J. L. Tallon, R. Michalak, and R. Dupree, Phys. Rev. B **54,** R 6909 (1996); K. Yamaya, T. Haga, and Y. Abe, J. Low Temp. Phys. **105**, 831 (1996).
[6]  Ch. Renner, B. Revaz, J. -Y. Genoud, K. Kadowaki, and O. Fischer, Phys. Rev. Letters **80**, 149 (1998); Ch. Renner, B. Revaz, J. -Y. Genoud, and O. Fischer, J. Low Temp. Phys*.* **105**, 1083 (1996).
[7]  B. Batlogg, H. Y. Hwang, H. Takagi, R. J. Cava, H. L. Kao, and J. Kwo, Physica C **235-240**, 130 (1994).
[8]  A. G. Loeser, Z. -X. Shen, D. S. Dessau, D. S. Marshall, C. H. Park, P. Fournier, and A. Kapitulnik, Science **273**, 325 (1996).
[9]  A. V. Chubukov and J. Schmalian, Phys. Rev. B**. 57,** R11085 (1998).
[10]  T. Ekino, T. Doukan, and H. Fujii, J. Low Temp. Phys. **105**, 563 (1996); J. F. Zasadzinski et al., J. Phys. Chem. Solids **53**, 1635 (1992).
[11]  G. V. M. Williams, J. L. Tallon, J. W. Quilty, H. J. Trodahl, and N. E. Flower, Phys. Rev. Letters **80**, 377 (1998).
[12]  J. W. Quilty, H. J. Trodahl, and D. M. Pooke, Phys. Rev. B **57**, R11097 (1998).
[13]  D. S. Marshall, D. S. Dessau, A. G. Loeser, C. -H. Park, A. Y. Matsuura, J. N. Eckstein, I. Bozovic, P. Fournier, A. Kapitulnik, W. E. Spicer, and Z. -X. Shen, Phys. Rev. Letters **76**, 4841 (1996).
[14]  G. V. M. Williams, J. L. Tallon, E. M. Haines, R. Michalak, and R. Dupree, Phys. Rev. Letters **78**, 721 (1997).
[15]  J. Ranninger and J. -M. Robin, Phys. Rev. B **56**, 8330 (1997).
[16]  C. P. Slichter, R. L. Corey, N. J. Curro, S. M. DeSoto, K. O'Hara, T. Imai, A. M. Kini, H. H. Wang, U. Geiser, J. M. Williams, K. Yoshimura, M. Katoh, and K. Kosuge, Phil. Mag. B **74**, 545 (1996).
[17]  N. Bulut and D. J. Scalapino, Phys. Rev. Letters **68**, 706 (1992).
[18]  K. A. Musaelian, J. Betouras, A. V. Chubukov, and R. Joynt, Phys. Rev. B **53**, 3598 (1996).

[19] J. K. Kirtley, C. C. Tsuei, J. Z. Sun, C. C. Chi, M. S. Yu - Jahnes, A. Gupta, M. Rupp, and I. M. B. Ketchen, Nature **373**, 225 (1995).
[20] M. Ido, M. Oda, N. Momono, C. Manabe, and T. Nakano, Physica C **263**, 225 (1996).
[21] D. Mandrus, J. Hartge, C. Kendziora, L. Mihaly, and L. Forro, Europhys. Letters **22**, 199 (1993); Y. DeWilde et al., Phys. Rev. Letters **80**, 153 (1998).
[22] T. Ekino, T. Minami, H. Fujii, and J. Akimitsu, Physica C **235-240**, 1899 (1994).
[23] M. Oda, K. Hoya, R. Kubota, C. Manabe, N. Momono, T. Nakano, and M. Ido, Physica C **282-287**, 1499 (1997).
[24] T. Hotta, J. Phys. Soc. Japan **62**, 274 (1993).
[25] Z. -X. Shen, W. E. Spicer, D. M. King, D. S. Dessau, and B. O. Wells, Science **267**, 343 (1995).
[26] J. Liu, Y. Li, and C. M. Lieber, Phys. Rev. B **49**, 6234 (1994).
[27] E. J. Nicol, H. Kim, M. Palumbo, and M. J. Graf, J. Low Temp. Phys. **105**, 539 (1996).
[28] J. H. Miller, Jr., Z. Zou, H. -M. Cho, J. Liu, Z. -S. Zheng, and W. -K. Chu, J. Low Temp. Phys. **105**, 527 (1996).
[29] C. Manabe, M. Oda, T. Nakano, and M. Ido, J. Low Temp. Phys. **105**, 489 (1996).
[30] Z. -X. Shen, D. S. Dessau, B. O. Wells, D. M. King, W. E. Spicer, A. J. Arko, D. Marshall, L. W. Lombardo, A. Kapitulnik, P. Dickinson, S. Doniach, J. DiCarlo, A. G. Loeser, and C. H. Park, Phys. Rev. Letters **70**, 1553 (1993).
[31] G. D. Mahan, Phys. Rev. Letters **71**, 4277 (1993); D. S. Dessau, Z. -X. Shen, and D. M. Marshall, Phys. Rev. Letters **71**, 4278 (1993).
[32] J. L. Tallon, C. Bernhard, U. Binninger, A. Hofer, G. V. M. Williams, E. J. Ansaldo, J. I. Budnick, and Ch. Niedermayer, Phys. Rev. Letters **74**, 1008 (1995).
[33] J. F. Annett, N. Goldenfeld, and A. J. Leggett, J. Low Temp. Phys. **105**, 473 (1996).
[34] C. T. Rieck, K. Scharnberg, S. Hensen, and G. Müller, J. Low Temp. Phys. **105**, 503 (1996); R. Hackl, M. Opel, P. F. Müller, G. Krug, B. Stadlober, R. Nemetschek, H. Berger, and L. Forró, J. Low Temp. Phys. **105**, 733 (1996).
[35] A. K. Rajagopal and S. S. Jha, in "Advances in Superconductivity -New Materials, Critical Currents and Devices", eds. R. Pinto, S. K. Malik, A. K. Grover, and P. Ayyub (New Age International, New Delhi, 1997), p. 378; A. K. Rajagopal and S. S. Jha, Phys. Rev. B **54**, 4331 (1996); A. Narlikar (ed.), "Studies of High Temperature Superconductors: Advances in Research & Applications, Vol. 1 (1989) – Vol. 8 (1991)" (Nova Science, New York).
[36] R. B. Laughlin, Phys. Rev. Letters **79**, 1726 (1997).
[37] P. A. Lee, J. Low Temp. Phys. **105**, 581 (1996).
[38] V. J. Emery and S. A. Kivelson, Nature **374**, 434 (1995).
[39] Z. -X. Shen and J. R. Schrieffer, Phys. Rev. Letters **78**, 1771 (1997).
[40] R. J. Radtke, K. Levin, H. -B. Schüttler, and M. R. Norman, Phys. Rev. B **48**, 15957 (1993).
[41] D. Brinkmann, Physica C **153-155**, 75 (1988); Y. Kitaoka, K. Ishida, K. Kondo, and K. Asayama, Physica B **165 & 166**, 1309 (1990); R. E. Walstedt, W. W. Warren, Jr., R. F. Bell, G. F. Brennert, G. P. Espinosa, J. P. Remeika, R. J. Cava, and E. A. Rietman, Phys. Rev. B. **36**, 5727 (1987); P. B. Allen and D. Rainer, Nature **349**, 396 (1991).
[42] A. J. Bray, M. A. Moore, and P. Reed, J. Phys. C : Solid St. Phys. **11**, 1187 (1978); T. A. Kaplan and N.d'Ambrumenil, ibid **15**, 3769 (1982); R. Rammal, R. Suchail, and R. Maynard, Solid State Commun. **32**, 487 (1979); D. Fiorani, J. Phys. C: Solid St. Phys. **17**, 4837 (1984); J. K. Srivastava, in "Selected Topics in Magnetism (Frontiers in Solid State Sciences, Vol. 2)", eds. L. C. Gupta and M. S. Multani (World Scientific, Singapore, 1993), p. 373.
[43] S. Morup, J. E. Knudsen, M. K. Nielsen, and G. Trumpy, J. Chem. Phys. **65**, 536 (1976); S. Morup, "Paramagnetic and Superparamagnetic Relaxation Phenomena Studied by Mössbauer Spectroscopy" (Polyteknisk Forlag, Lyngby, 1981); B. V. Thosar, P. K. Iyengar, J. K. Srivastava and S. C. Bhargava (eds.), "Advances in Mössbauer Spectroscopy: Applications to Physics, Chemistry and biology" (Elsevier, Amsterdam, 1983).
[44] Ch. Renner and O. Fischer, Phys. Rev. B **51**, 9208 (1995).
[45] G. Burns, "High Temperature Superconductivity – An Introduction" (Academic, San Diego, 1992).
[46] L. Solymar, "Superconductive Tunnelling and Applications" (Chapman and Hall, London, 1972).
[47] E. L. Wolf, "Principles of Electron Tunneling Spectroscopy" (Oxford Univ. Press, New York, 1985).
[48] R. C. Dynes, J. P. Garno, G. B. Hertel, and T. P. Orlando, Phys. Rev. Letters **53**, 2437 (1984).
[49] S. M. Freake and C. J. Adkins, Phys. Letters **29A**, 382 (1969).

- [50] T. Claeson and S. Gygax, Solid State Commun. **4**, 385 (1966).
- [51] T. R. Lemberger and L. Coffey, Phys. Rev. B **38**, 7058 (1988).
- [52] A. S. Alexandrov and N. F. Mott, "High Temperature Superconductors and Other Superfluids" (Taylor & Francis, London, 1994).
- [53] Y. Kitaoka, H. Yamada, K. Ueda, Y. Kohori, T. Kohara, Y. Oda, and K. Asayama, Japan. J. Appl. Phys. **26** (Suppl. 26-3), 1221 (1987); J. H. Kang, J. Maps, and A. M. Goldman, ibid p. 1233; Y. Kohori, T. Kohara, H. Shibai, Y. Oda, T. Kaneko, Y. Kitaoka, and K. Asayama, ibid p. 1239; Y. Xu and R. N. Shelton, ibid p. 1269.
- [54] T. Shibauchi, L. Krusin-Elbaum, Ming Li, M. P. Maley, and P. H. Kes, Phys. Rev. Letters **86**, 5763 (2001).
- [55] J. K. Srivastava, cond-mat/0503711 (March 2005).